\documentclass[twocolumn,showpacs,preprintnumbers,amsmath,amssymb]{revtex4}

\usepackage{amsfonts}
\usepackage{graphicx}
\usepackage{dcolumn}
\usepackage{bm}
\usepackage{balance}
\usepackage{amsmath}

\begin{document}

\title{Multipartite entanglement serves as a faithful detector for quantum phase transitions }

\author{ Yan-Chao Li }
\email{ycli@ucas.ac.cn} \affiliation{ \textit Center of Materials
Science and Optoelectronics Engineering, College of Materials
Science and Opto-Electronic Technology, University of Chinese
Academy of Sciences, Beijing 100049, China}
\author{ Yuan-Hang Zhou }
\author{Yuan Zhang} \affiliation{ \textit Center of Materials Science
and Optoelectronics Engineering, College of Materials Science and
Opto-Electronic Technology, University of Chinese Academy of
Sciences, Beijing 100049, China}

\author{ Yan-Kui Bai }
\affiliation{ \textit  College of Physics and Hebei Key Laboratory
of Photophysics Research and Application, Hebei Normal University,
Shijiazhuang, Hebei 050024, China}

\author{ Hai-Qing Lin }
\email{hqlin@zju.edu.cn}\affiliation{\textit Zhejiang Institute of
Modern Physics and School of Physics, Zhejiang University, Hangzhou
310027, China}

\date{ \today }

\begin{abstract}

We investigate quantum phase transitions in various spin chain
systems using the multipartite entanglement measure $\tau_{SEF}$
based on the monogamy inequality of squared entanglement of
formation. Our results demonstrate that $\tau_{SEF}$ is more
effective and reliable than bipartite entanglement or bipartite
correlation measures such as entanglement of formation, von Neumann
entropy, and quantum discord in characterizing quantum phase
transitions. $\tau_{SEF}$ not only detects critical points that may
go unnoticed by other detectors but also avoids the issue of
singularity at non-critical points encountered by other measures.
Furthermore, by applying $\tau_{SEF}$, we have obtained the phase
diagram for the XY spin chain with three and four interactions and
discovered a new quantum phase.

\end{abstract}

\maketitle

\section{\label{Sec:level1} Introduction}
Entanglement is a pure quantum phenomenon and an essential feature
that distinguishes quantum mechanics from classical
mechanics~\cite{Horodecki09}. Its nonlocality has always puzzled
scientists~\cite{Einstein35,Bell87,Hardy93,Loss00}, and essentially
all significant experiments to date have supported the
quantum-mechanical
predictions~\cite{Bell64,Freedman72,Aspect82,Weihs98,Kofler16,Handsteiner17}.
On the other hand, entanglement can serve as a carrier of quantum
information and is considered a crucial resource in quantum
information processing. Therefore, the study of entanglement has
always held a significant position in the field of quantum
information~\cite{Shor00,Lo03,Nielsen00,Amico08}.

Quantum phase transitions (QPTs)~\cite{Sachdev99}, which are purely
driven by quantum fluctuations, occupy a significant position in
quantum many-body systems. Researchers even believe that QPTs play
significant roles in understanding many puzzles in physics, such as
heavy-fermion metals and high-temperature
superconductors~\cite{Gegenwart08}.

Given the inherited quantum nature, Osterloh \textit{et
al.}~\cite{Osterloh02} first introduced the entanglement concept to
study QPTs successfully. Compared to the traditional way of
Landau-Ginzburg-Wilson spontaneous symmetry breaking theory, no a
priori knowledge of symmetry of the system is needed for the new
method. Subsequently, many detectors borrowed from quantum
information science are used to characterize quantum criticality.
These include the von Neumann
entropy~\cite{Gu04,Legeza06},entanglement of formation
(EOF)~\cite{Gu042,LYC16,Xu23}, quantum
fidelity~\cite{Zanardi06,Quan06}, quantum discord
(QD)~\cite{Werlang10,LYC11}, quantum coherence (QC) based on Wigner
and Yanase skewed information~\cite{Karpat14,LYC20,Lv22}, steered
quantum coherence~\cite{Hu20}, and magic resource~\cite{Fu22,Fu23}.
However, these proposed detectors have limitations in detecting
certain types of QPTs~\cite{Buonsante07,Chen08,LYC20,Lv22}, such as
in detecting the Berezinskii-Kosterlitz-Thouless (BKT)-type QPT in
the XXZ model. Finding an effective and universally applicable
detector for all kinds of QPTs has always been a goal of
researchers~\cite{Legeza06,Buonsante07,Hu20,Nandi22}.

Many-body systems often exhibit multipartite entanglement or
correlation (it is possible for more than two parties to be
entangled in different inequivalent ways, and multipartite
entanglement can inevitably reflect a greater amount of diverse and
sophisticated quantum information~\cite{Chiara18}). Therefore, it
should obviously have an advantage in characterizing quantum phase
transitions, as reflected by changes in the degree of entanglement.
However, most of the QPT detectors currently being studied fall into
the category of bipartite correlation. Although there have been some
studies on the relationship between multipartite entanglement and
quantum phase transitions~\cite{Oliveira06,Xu17,Xu18,Xu21,Su22},
most of them focus on identifying specific quantum phase transitions
and there is still lack of research on the universality of the
ability in detecting QPTs of multipartite entanglement. Therefore,
in this paper, we focus on the effectiveness of multipartite
entanglement in determining quantum phase transitions for different
systems. The corresponding quantum phase transitions, which vary
depending on the system being studied, include the first-order type
and BKT-type QPTs observed in the XXZ model, as well as the
topological-type QPT seen in the SSH model, among others. We also
explore the characteristics and advantages of using multipartite
entanglement compared to bipartite entanglement or bipartite
correlation detectors for determining quantum phase transitions.

\section{\label{sec:level2} Methods}
Entanglement monogamy is one of the most important properties in
multipartite systems, which implies that quantum entanglement cannot
be freely shared in many-body systems~\cite{ben96pra}. Based on
quantitative entanglement monogamy inequalities
\cite{ckw00pra,fan07pra,Bai13,Bai14prl}, one can construct effective
measures or indicators for multipartite entanglement. The
entanglement of formation possesses operational physical meaning and
is related to quantum data compression \cite{Wootters98}. It has
been proven that the squared entanglement of formation obeys the
monogamy inequality and its residual entanglement has been
demonstrated to be a trustworthy measure for genuine multipartite
entanglement in quantum many-body systems~\cite{Bai14prl,Bai14pra}.

In an $N$-qubit multipartite system, genuine multipartite
entanglement based on the residual entanglement of formation can be
expressed as \cite{Bai14prl}
\begin{eqnarray}\label{eq:1}
\tau_{SEF}=E_f^2\left(\rho_{A_1|A_2\cdots
A_N}\right)-\sum_{k\neq1}^N E_f^2\left( \rho_{A_1A_k}\right),
\end{eqnarray}
where $E_f\left(\rho_{A_1|A_2\cdots A_N}\right)$ quantifies the
entanglement between one arbitrary selected qubit $A_1$ and the rest
qubits of the system, and can be written as
$E\left(\rho_{A_1|A_2\cdots
A_N}\right)=min\Sigma_sp_sE^v_1(\rho_{A_1}^s)$, in which
$E^v_1(\rho_{A_1}^s)=-Tr\rho_{A_1}^s log_2\rho_{A_1}^s$ is the von
Neumann entropy derived from the single-qubit reduced density matrix
$\rho_{A_1}$ and the minimum runs over all the pure state
decompositions $\rho^s=\Sigma_sp_s\left| \psi_{A_1\cdots A_N}^s
\right\rangle \left\langle \psi_{A_1\cdots
A_N}^s\right|$~\cite{Bennett96,Su22} (if the state considered is
itself a pure state, there is no need to perform the decomposition
as we do in this paper); $E_f\left( \rho_{A_1A_k}\right)$ is the
entanglement of formation (EOF) between the two qubits $A_1$ and
$A_k$, and can be defined as~\cite{Werlang10,Maziero10}
\begin{align}
EOF(\rho_{A_1A_k})=&-x\log_2x
-\left[1-x\right]\log_2\left[1-x\right],\label{eq:2}
\end{align}
where $x=(1+\sqrt{1-C_{\rho_{A_1A_k}}^2})/2$,
$C_{\rho_{A_1A_k}}\equiv\max\{0,\lambda_1-\lambda_2-\lambda_3-\lambda_4\}$
is the concurrence~\cite{Wootters98}, where $\lambda_n$ ($n=1,2,3,$
and $4$) are the square roots of the eigenvalues of
$\rho_{A_1A_k}\tilde{\rho}_{A_1A_k}$ in descending order, and
$\tilde\rho_{A_1A_k}=(\sigma_{A_1}^y\otimes\sigma_{A_k}^y)\rho_{A_1A_k}^\ast(\sigma_{A_1}^y\otimes\sigma_{A_k}^y)$
represents the time-reversed matrix of $\rho_{A_1A_k}$,
$\rho_{A_1A_k}^\ast$ is the complex conjugation of $\rho_{A_1A_k}$,
and $\sigma^y$ refers to the $y$ component of the Pauli operator.

In addition to EOF, for a comprehensive comparison, we also consider
other bipartite correlation detectors that are currently more
effective in determining quantum phase transitions: quantum discord
and quantum entanglement based on von Neumann entropy. Here, we
consider the two-site entanglement entropy ~\cite{Gu04}
\begin{align}\label{eq:3}
E^v_2(\rho_{r}) =-\texttt{Tr}\rho_{r}\texttt{ln}\rho_{r},
\end{align}
where $\rho_{r}$ represents the two-site reduced density matrix of
$A_1$ and $A_j$, which are two lattice qubits separated by a
distance $r=j-1$. Compared to the single-site entropy used in
Eq.(\ref{eq:1}), $E_2^v(\rho_{r})$ only include one more qubit in
the reduced density matrix used for calculation, but is more
effective in detecting QPTs~\cite{Legeza06}. The QD can be written
as
\begin{align}\label{eq:4}
QD(\rho_{A_1A_j})=&E^v_1(\rho_{A_j})+\min_{\{A_{j_k}\}}\widetilde{E_v}(\rho_{A_1A_j}|\{A_{j_k}\})\nonumber\\
&-E^v_2(\rho_{A_1A_j}),
\end{align}
where
${\widetilde{E_v}(\rho_{A_1A_j}|\{A_{j_k}\})}=\sum_kp_kE^v_2(\rho_{A_1A_j}^k)$
with $\rho_{A_1A_j}^k=\frac{1}{p_k}$ $\left(I\otimes
A_{j_k}\right)\rho_{A_1A_j}\left(I\otimes A_{j_k}\right)$ and
$p_k=\texttt{Tr}$ $\left[\left(I\otimes
A_{j_k}\right)\rho_{A_1A_j}\left(I\otimes A_{j_k}\right)\right]$ is
the conditional entropy, and the minimum is achieved from a complete
set of projective measures $\left\{A_{j_k}\right\}$ on site
$A_j$~\cite{Ollivier01,Dillenschneider08,Werlang10}. The projectors
here can be written as
\begin{align}\label{eq:5}
A_{j_k}=V\left|k\right\rangle\left\langle k\right|V^\dagger,
\end{align}
where $\{\left|k\right\rangle\}$ is the standard basis
$\{\left|\uparrow\right\rangle,\left|\downarrow\right\rangle\}$ of
any two selected spins, and the transform matrix $V$ is
parameterized as~\cite{Sarandy09}
\begin{align}\label{eq:6}
V=\left(
  \begin{array}{cccc}
    \cos\theta & e^{-i\varphi}\sin\theta \\
    e^{i\varphi}\sin\theta  &  -\cos\theta \\
    \end{array}
\right).
\end{align}
Then the minimum of the conditional entropy
$\widetilde{S}(\rho_{r})$ can be obtained by traversing the $\theta$
and $\varphi$.

To calculate the above detectors, we use the exact diagonalization
(ED) techniques to simulate the systems. To strictly construct the
matrix form of the Hamiltonian, we choose the $S_z$ representation
and work in the whole $S_z$ space. Periodic boundary conditions are
considered for all the calculated systems to reduce the influence of
the boundary. In addition, we point out that, to calculate
$\tau_SEF$, it is necessary to calculate the EOF between the
selected qubit and all other qubits. This is undoubtedly a challenge
for large systems, but we found that due to the periodic boundary
conditions, the EOF between corresponding lattice sites located on
either side of the selected lattice site has a symmetric
characteristic. Therefore, only the situation on one side needs to
be calculated. Moreover, it was found that the formation of
entanglement rapidly decreases with the increase of qubits spacing.
Therefore, in practical calculations, it is usually only necessary
to calculate some lattice sites with close distances. This has
important practical significance for large-scale systems, such as
the possible future consideration of using the density matrix
renormalization group (DMRG) numerical calculation method to
calculate larger systems.

\begin{figure}[t]
\includegraphics[width=1.0\columnwidth]{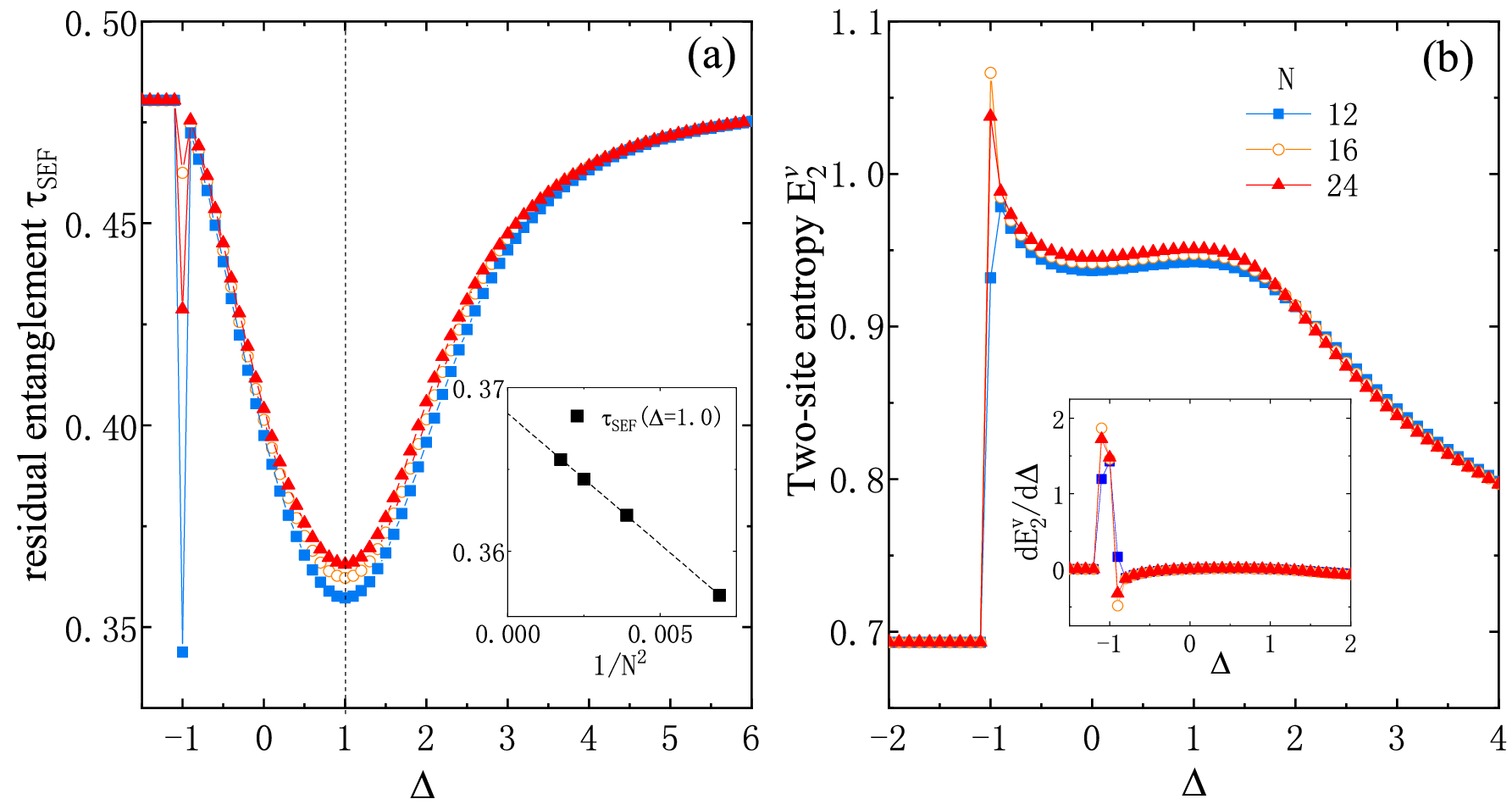}
\caption{\label{fig1} (Color online) (a) Residual entanglement
$\tau_{SEF}$ and (b) two-site entropy $E_2^v$ as functions of
$\Delta$ for the XXZ model, with different system sizes $N$. The
inset in (a) shows the finite size-scaling behavior of the minimum
value of $\tau_{SEF}$ at $\delta=1.0$.}
\end{figure}

\section{\label{sec:level3}Results and discussions}
\subsection*{\label{sec:level3a} A. The BKT-type
QPT in the XXZ model}
The Hamiltonian for the XXZ spin model is defined as follows:
\begin{eqnarray}\label{eq:7} H_{XXZ}=\sum_j^N\sigma _{j}^{x}\sigma
_{j+1}^{x}+\sigma _{j}^{y}\sigma _{j+1}^{y}+\Delta \sigma
_{j}^{z}\sigma _{j+1}^{z},
\end{eqnarray}
where ${\Delta}$ describes the anisotropy of the spin-spin
interaction in the z direction, and $\sigma_j$ represents the usual
Pauli matrices at site $j$. It is well known that the model is in a
critical phase for $-1\leq\Delta\leq1$, an antiferromagnetic phase
at $\Delta\geq1$, and ferromagnetic phase at $\Delta<-1$. The
critical point (CP) at $\Delta=1$ belongs to the continuous BKT
type, while the CP at $\Delta=-1$ is a fist-order transition caused
by the ground state level crossing~\cite{Takahashi99}.

The $\tau_{SEF}$ and $E^v_2$ results from the exact diagonalization
for different system sizes $N$ with periodic boundary conditions are
shown in Fig.~\ref{fig1}. The first order CP at $\Delta=-1$ is
clearly detected by both detectors, while the BKT-type CP at
$\Delta=1$ can only be reflected by the minimum of $\tau_{SEF}$ (see
Fig.~\ref{fig1}(a)). The location of the minimum of $\tau_{SEF}$
does not change with $N$, and it will exist even in the
thermodynamic limit: the value of the minimum is linearly related to
$1/N^2$. As $N$ tends to infinity, meaning that $1/N^2$ tends to 0,
it approaches a fixed value of 0.36(8) (see the size scaling
behavior in the inset of Fig.~\ref{fig1}(a)). However, for this CP,
$E^v_2$ exhibits a smooth curve behavior, as shown in
Fig.~\ref{fig1}(b), and its derivative does not exhibit any
singularity (see the inset of Fig.~\ref{fig1}(b)); thus, it cannot
reflect the BKT-type transition. This result is in agreement with
the results of the TMRG in Ref.~\cite{Lv22}.

We must emphasize that both quantum discord (QD) and entanglement of
formation (EOF) can indeed reflect phase transitions in the XXZ
model, and the corresponding findings have been published in
Refs.~\cite{Lv22,Werlang2010}. However, our primary objective was to
demonstrate the broader applicability of residual entanglement as a
detector for quantum phase transitions compared to other detection
methods. Therefore, our focus was on examining whether residual
entanglement can overcome the limitations of other detectors. Given
that enumerating the corresponding results for QD and EOF would not
align with our main theme, we opted not to include them here.

\begin{figure}[b]
\includegraphics[width=1.0\columnwidth]{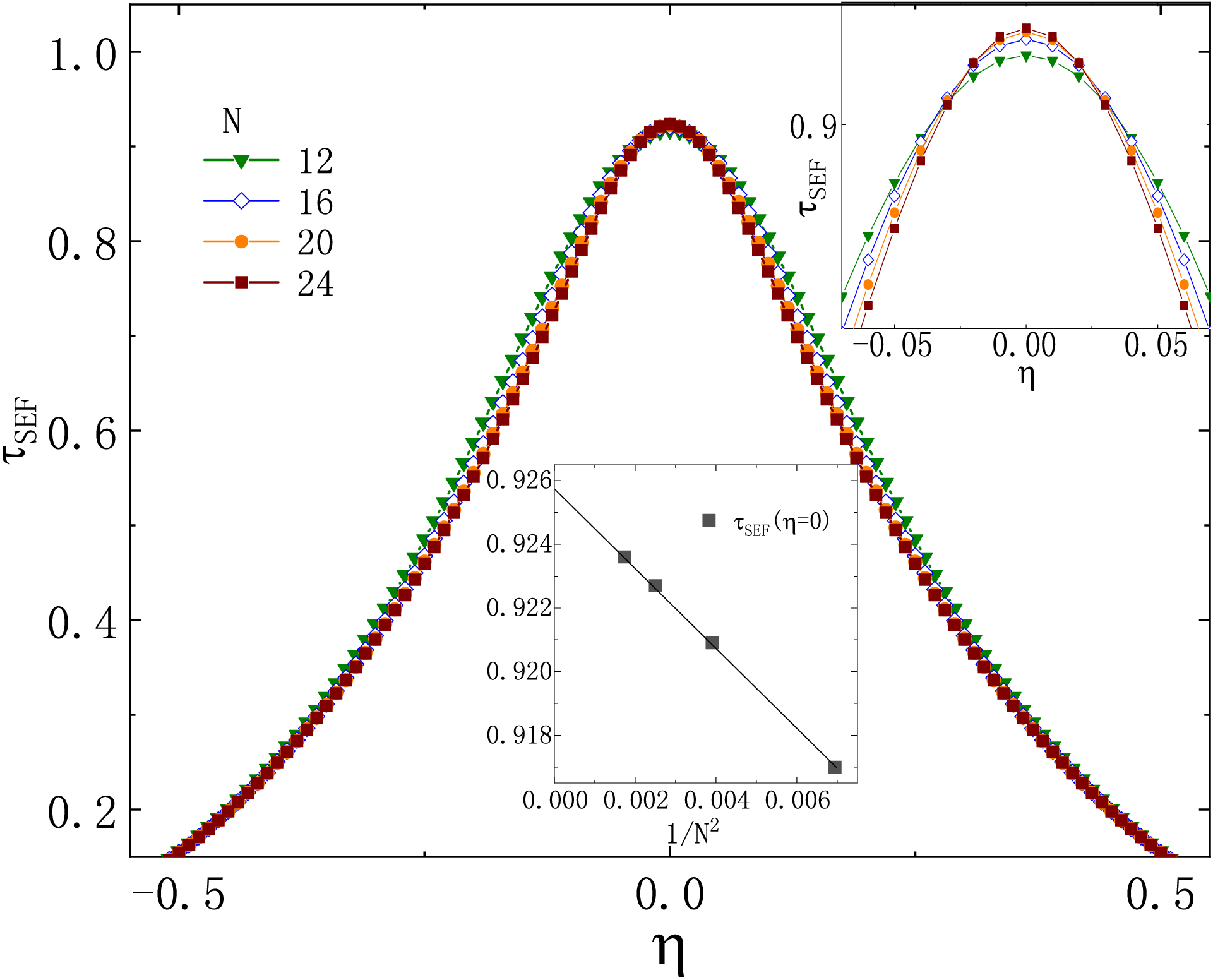}
\caption{\label{fig2} (Color online) Residual entanglement
$\tau_{SEF}$ as a function of $\eta$ under different system sizes
$N$ for the SSH model. The inset at the bottom of the figure
illustrates the finite-size scaling behavior of the maximum value of
$\tau_{SEF}$ at $\eta=0$, while the inset in the upper right corner
of the figure provides an enlarged view of the curves of
$\tau_{SEF}$ near $\eta=0$ for better clarity.}
\end{figure}

\subsection*{\label{sec:level3b} B. The topological QPT in the SSH model}
Then we consider the topological QPT in the one-dimensional
Su-Schrieffer-Heeger (SSH) model~\cite{Wakatsuki14}. The Hamiltonian
is determined as follows: \begin{align}\label{eq:8}
H=&-\sum_{j}\left[(1+\eta)c_{j,B}^\dag
c_{j,A}+(1-\eta)c_{j+1,A}^\dag c_{j,B}+H.c.\right],
\end{align}
where the operator $c_{j,\alpha}$ destroys a spinless fermion at the
unit cell $j$ of type $\alpha = A, B$ (where $A$ and $B$ are the
sublattice indices), and $\eta$ represents the dimerization. There
is a topological phase transition at $\eta=0$. When $\eta$
transitions from being negative to positive, the system undergoes a
transformation from a topological phase to a topological trivial
phase~\cite{Wakatsuki14,Yu16}. Different from the general QPTs, a
topological phase transition can not be characterized by a local
order parameter~\cite{Sachdev99,Wen04}.

\begin{figure}[t]
\includegraphics[width=0.8\columnwidth]{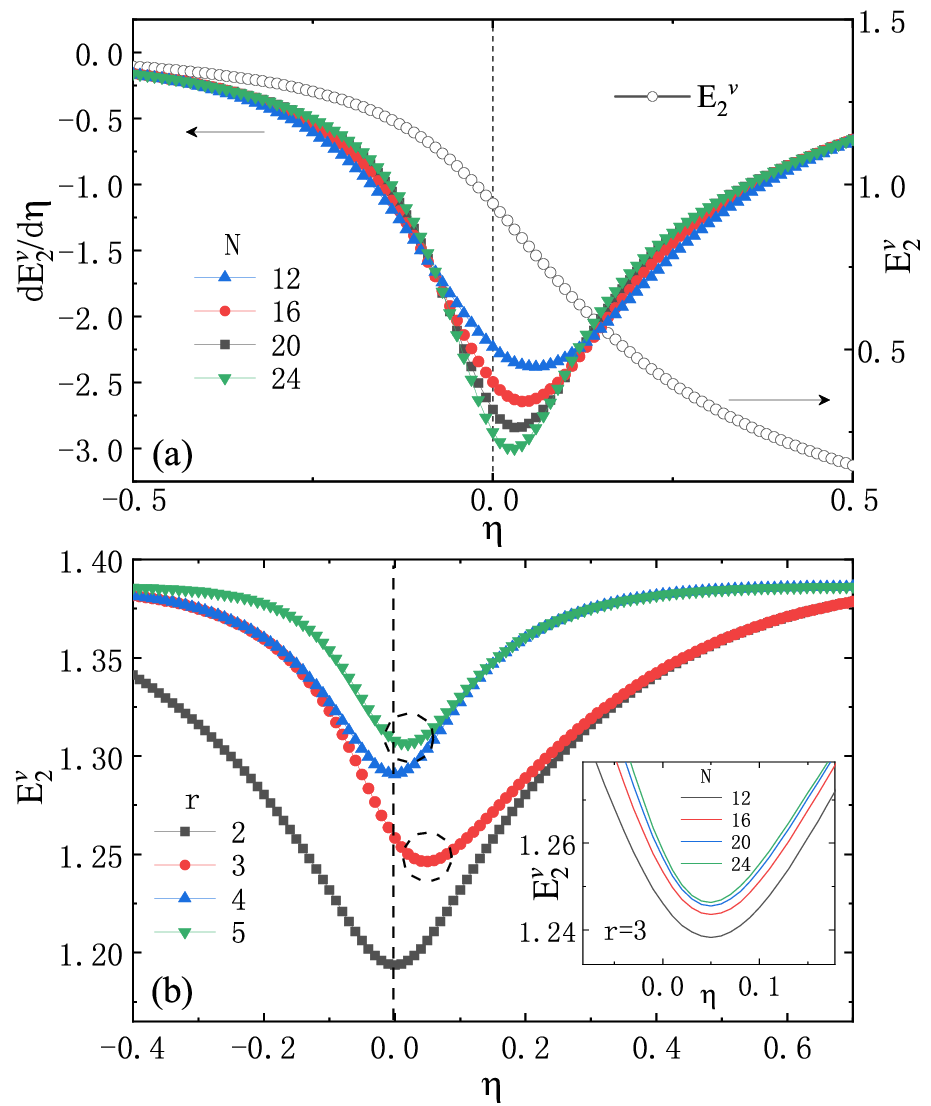}
\caption{\label{fig3} (Color online) (a) $E_2^v$ and $dE_2^v/d\eta$
as a function of $\eta$ for $r=1$ and different N in the SSH model.
The arrows indicate the corresponding vertical axis of these curves.
(b) $E_2^v$ as a function of $\eta$ for different values of $r$ when
$N=24$. The dashed circles indicate the minimum points of the
curves, and the inset demonstrates that the valley position for
$r=3$ is less affected by the system size $N$.}
\end{figure}

\begin{figure}[t]
\includegraphics[width=1.0\columnwidth]{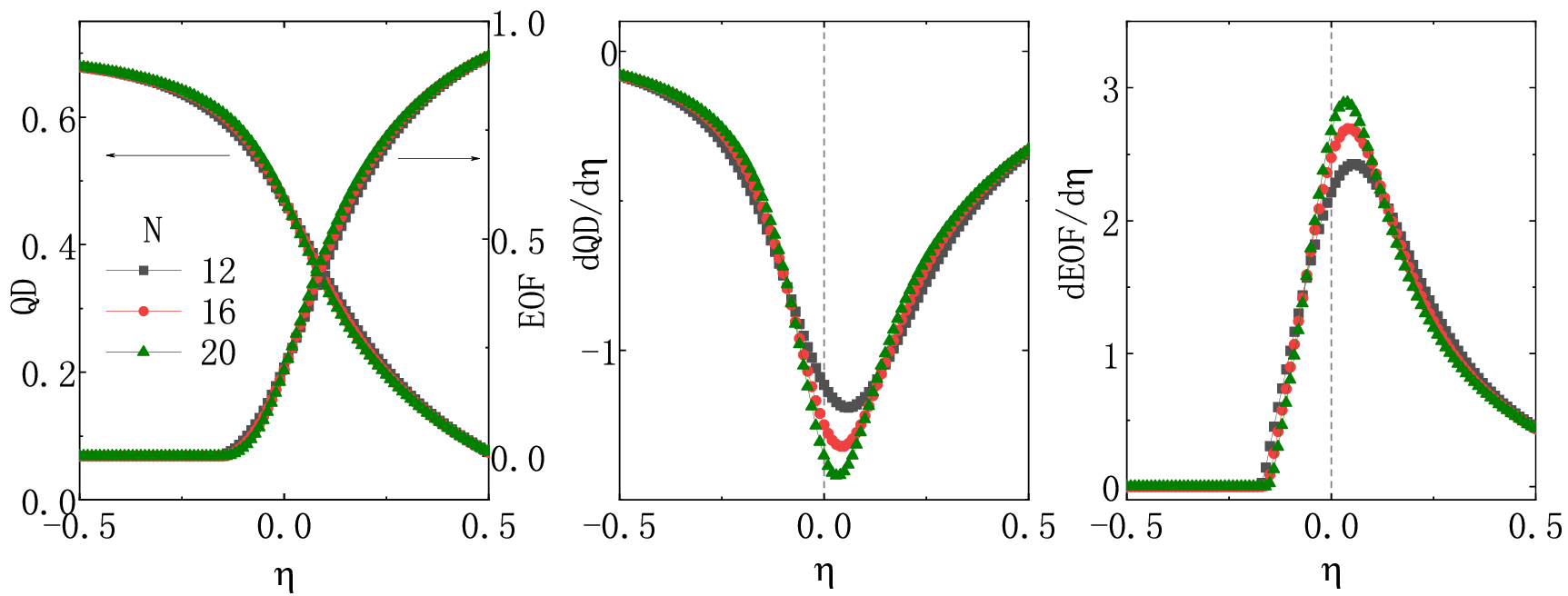}
\caption{\label{fig32} (Color online) (a) $QD$ and $EOF$ as a
function of $\eta$ for $r=1$ under different N for the SSH model.
The arrows indicate the corresponding vertical axis of these curves.
(b) and (c) show the derivatives of QD and EOF in (a), respectively,
where the peak or deep position does not directly locate at the
accurate QPT point. In order to obtain the accurate critical point,
we need to conduct the finite-size scaling analysis. }
\end{figure}

The multipartite entanglement results of $\tau_{SEF}$ are shown in
Fig.~\ref{fig2}. There is a maximum that appears at $\eta=0$, and it
becomes clearer as $N$ increases. The value of $\tau_{SEF}$ at $\eta
= 0$ increases linearly with decreasing $1/N^2$, reaching a maximum
value of approximately $0.925(8)$ when $N$ approaches infinity,
specifically when $1/N^2 = 0$, as shown in the inset of
Fig.~\ref{fig2}. The maximum clearly indicates the topological QPT
of the system. As a comparison, the bipartite entanglement entropy
$E_2^v$ changes continuously and smoothly near $\eta=0$, which does
not reflect the QPT. To reflect its occurrence, we need to use the
derivative $dE_2^v/d\eta$, which exhibits a minimum behavior, and
the minimum position $\eta_m$ can be considered as a pseudo-critical
point~\cite{Barbar83,Zhu06}: as the system size $N$ increases, the
minimum becomes more pronounced and moves towards the critical point
$\eta_c=0$ (see Fig.~\ref{fig3}). The performance of the other
detectors, such as the QD and EOF, show similar behaviors (see
Fig.~\ref{fig32}). In order to obtain the accurate critical point,
we need to conduct the finite-size scaling analysis. However,
scaling analysis of most practical systems at large scales is often
difficult. Therefore, it has certain advantages in determining
quantum phase transitions for the behavior of the residual
entanglement $\tau_{SEF}$, where the indicated critical point is
independent of the system size.

Further research finds that when the spacing between the two
selected spins $r>1$, $E_2^v$ can also exhibit a minimum behavior
near the phase transition point $\eta_c$, as shown in
Fig.~\ref{fig3}(b). When $r$ is even, the minimum point accurately
corresponds to $\eta_c$. When $r$ is odd, the position of the
minimum significantly deviates from $\eta_c$ --- the smaller $r$ is,
the greater the deviation. Moreover, this deviation does not change
significantly as $N$ increases, as shown in the inset of
Fig.~\ref{fig3}(b). Consequently, to obtain the exact location of
the phase transition point, it may be necessary to calculate on a
significantly larger scale $N$ or consider $r=N/2$, which represents
the maximum distance between two spins in the system. We believe
that the reason for this is related to the structure of the
spin-spin interaction in the system. The variation in the
interaction of odd and even bonds gives rise to different symmetries
in the bond interactions between the selected pair of spins and the
remaining spins. If it is symmetrical, it will align with the
overall symmetry of the system's ground-state vector. Conversely, a
larger size is necessary to minimize the influence of the selected
spin positions on the overall symmetry.

\subsection*{\label{sec:level3c} C. The QPT in the SSH-XY model}
The quantum phase transitions mentioned above are well known to
researchers. Next, we consider an unknown quantum phase transition
that occurs in a novel system. We call this model the SSH-XY model,
which is similar to the SSH-XXZ model in Ref.~\cite{LYC19}, and its
Hamiltonian can be written as
\begin{eqnarray}\label{eq:9}
H&=&-\sum_{j=1}^N\left(\frac{1+\gamma_1}{2}\sigma_{2j-1}^x\sigma_{2j}^x+\frac{1-\gamma_1}{2}\sigma_{2j-1}^y\sigma_{2j}^y\right)\nonumber\\
&+&\left(\frac{1+\gamma_2}{2}\sigma_{2j}^x\sigma_{2j+1}^x+\frac{1-\gamma_2}{2}\sigma_{2j}^y\sigma_{2j+1}^y\right),
\end{eqnarray}
where ${\sigma^\alpha}$ represents the Pauli matrices
(${\alpha=x,y}$) as those in Eq.(\ref{eq:7}).${\gamma_1}$ and
${\gamma_2}$ describe the anisotropy in the XY directions that
arises from the spin-spin interaction between the odd and even
bonds, respectively.

\begin{figure}[t]
\includegraphics[width=7cm]{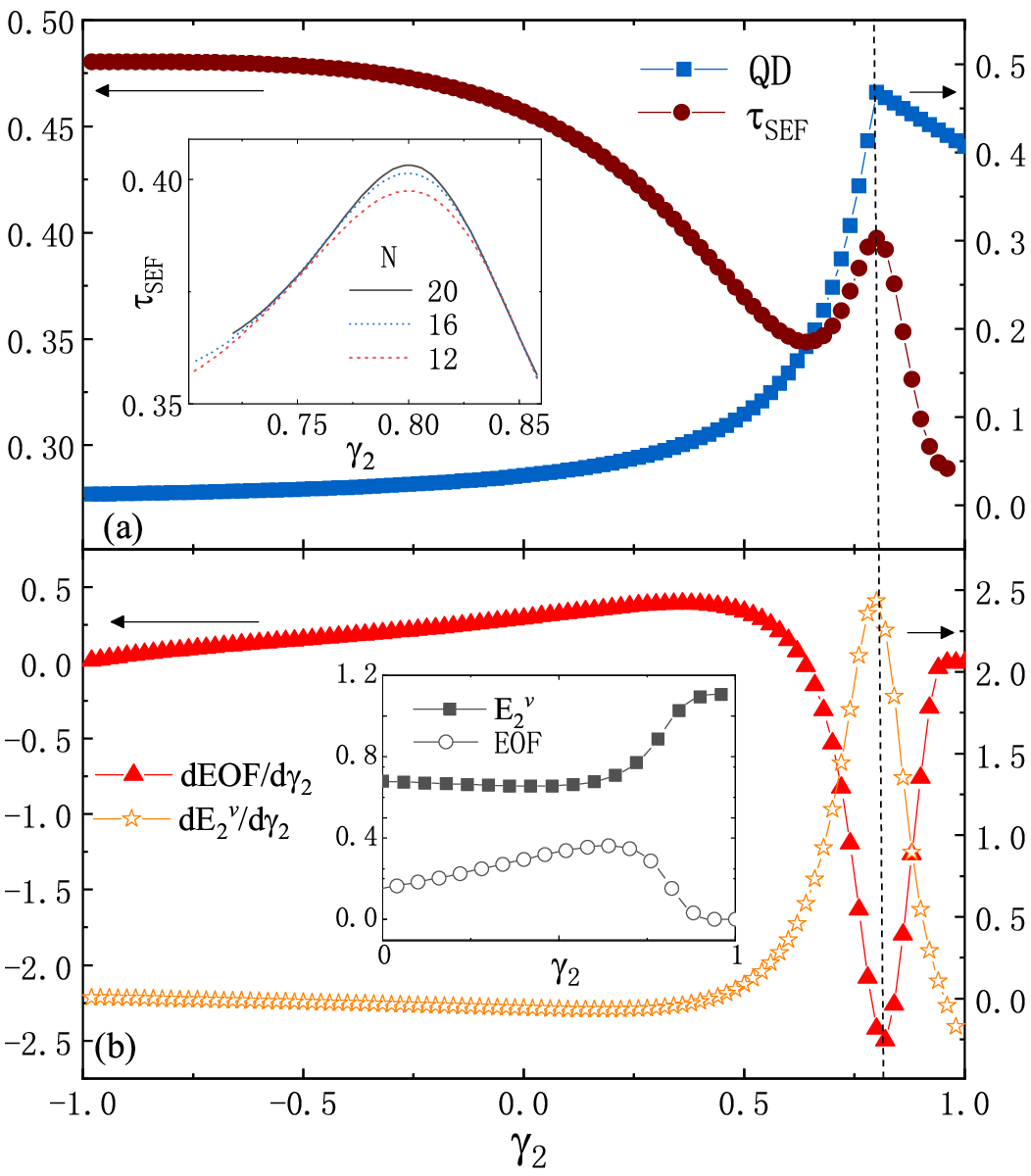}
\caption{\label{fig4} (Color online) (a) QD and residual
entanglement $\tau_{SEF}$ and (b) the $\gamma_2$ derivatives of EOF
and $E_2^v$ as a function of $\gamma_2$ under $\gamma_1=-0.8$ and
$N=20$. The inset in (a) shows the peak portion of $\tau_{SEF}$ for
different values of $N$, while the behaviors of EOF and $E_2^v$ are
displayed as an inset in (b). The arrows indicate the vertical axis
of the corresponding curves. }
\end{figure}

The results for $\gamma_1=-0.8$ are plotted in Fig.~\ref{fig4}.
$\tau_{SEF}$ displays a sharp peak behavior at $\gamma_2=0.8$, which
indicates a quantum phase transition. To further demonstrate this,
we also calculated other detectors, as shown in Fig.~\ref{fig4}. The
QD displays a non-analytical turning point at the same parameter
position, whereas the EOF and von Neumann entropy $E_2^v$ do not
exhibit singularities (see the inset in Fig.~\ref{fig4}(b)) but
their first-order derivatives exhibit extreme behavior at this
location. As the value of $N$ increases, these behaviors become
increasingly evident (the $\tau_{SEF}$ results for different values
of $N$ are presented in the inset in Fig.~\ref{fig4}(a)), indicating
the existence of a quantum phase transition point at $\gamma_2=0.8$.
QD has similar results, and for EOF and $E_2^v$, regardless of how
large $N$ is, the derivative behavior is necessary to capture the
phase transition, so only the results for $N=20$ are given here. In
addition, the deep behavior near at $\gamma_2=0.66$ for $\tau_{SEF}$
is weakened as $N$ increases, tending towards the critical point
$\gamma_2=0.8$ and thus, is not an indicator of QPT.

The occurrence of this phase transition is actually easy to
understand. It is the result of the competition between the $X$ and
$Y$ directions from the odd and even coupling bonds. In the process
where $\gamma_2$ is constantly increasing from $-1$, the
corresponding system will experience an increase in coupling in the
$X$ direction and a decrease in coupling in the $Y$ direction. When
$\gamma_2>0.8$, the coupling in the $X$ direction on the even bond
dominates the system, resulting in a phase transition similar to the
phase transition point at $\gamma=0$ in the XY model. The results
remain completely consistent for other values of $\gamma_1$, and we
have chosen $\gamma_1=-0.8$ as a representative point for
demonstration purposes only. Here, $\tau_{SEF}$, as well as QD,
which contains more quantum correlation information compared to
two-body entanglement, can accurately reflect the occurrence of
phase transitions without requiring derivation.

\subsection*{\label{sec:level3d} D. XY model with multi-site interactions(XYMI)}
The Hamiltonian of the $XY$ spin model in a transverse field with
three- and four-spin interactions can be written as follows:
\begin{align}\label{eq:10}
H=&-\sum_{j=1}^N[(\frac{1+\gamma}{2}\sigma_j^x\sigma_{j+1}^x+\frac{1-\gamma}{2}\sigma_j^y\sigma_{j+1}^y+
\lambda\sigma_j^z)\nonumber\\
&+\alpha\left(\sigma_{j-1}^x\sigma_j^z\sigma_{j+1}^x+\sigma_{j-1}^y\sigma_j^z\sigma_{j+1}^y\right)\nonumber\\
&\left.+\beta\left(\sigma_{j-1}^x\sigma_j^z\sigma_{j+1}^z\sigma_{j+2}^x+\sigma_{j-1}^y\sigma_j^z\sigma_{j+1}^z\sigma_{j+2}^y\right)\right],
\end{align}
where ${\sigma^\alpha}$, ${\gamma}$, and ${\lambda}$ have the same
meaning as those in Eq.(\ref{eq:7}); ${\alpha}$ and $\beta$ denote
the three-spin and fourth-spin interactions, respectively.
\begin{figure}[t]
\includegraphics[width=1.0\columnwidth]{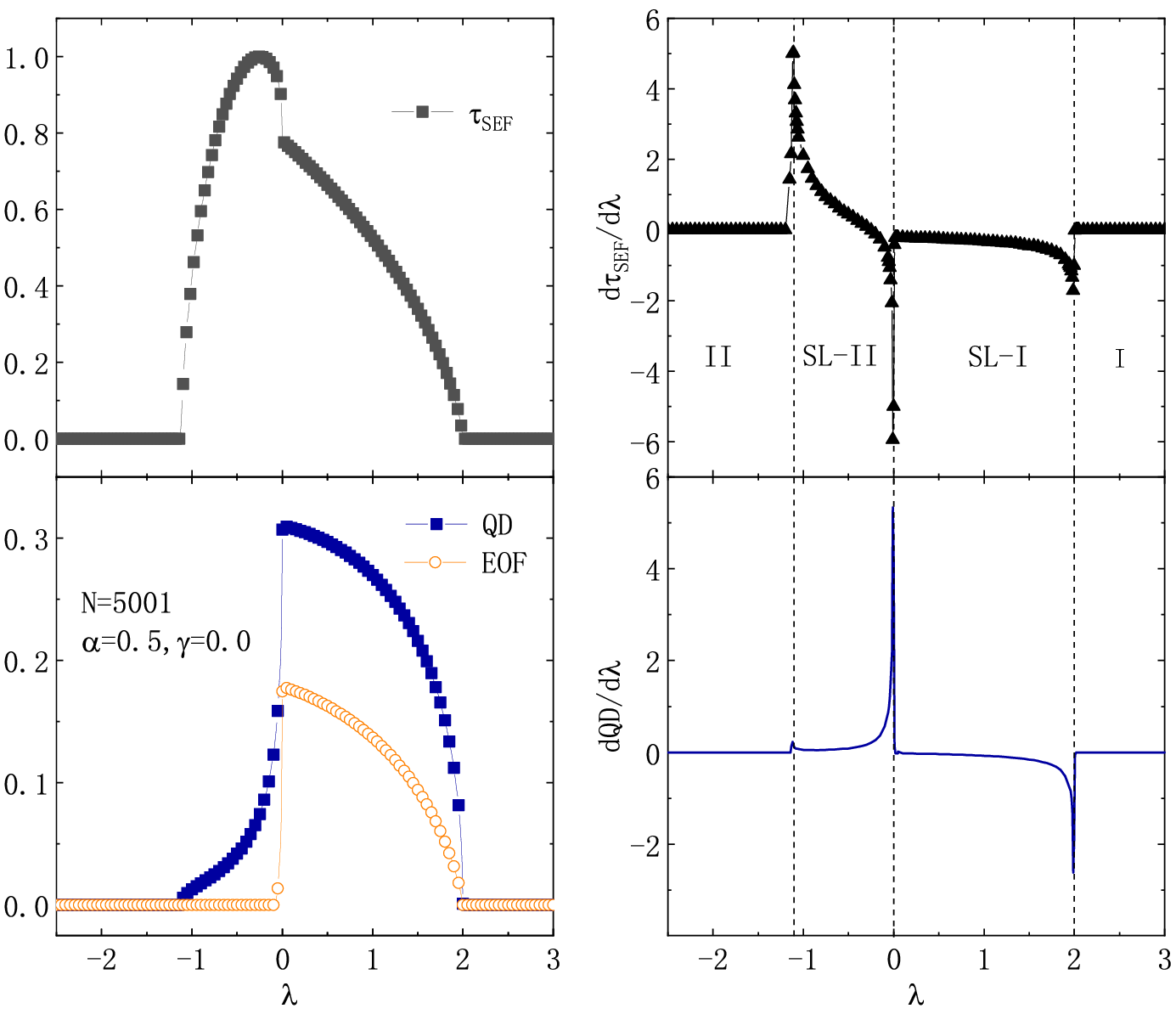}
\caption{\label{fig5} (Color online) (a) $\tau_{SEF}$ and (b) QD and
EOF as functions of $\lambda$ under $\gamma=0$, $\alpha=0.5$,
$\beta=0$, and $N=5001$ for the XYMI model. The derivatives of
$\tau_{SEF}$ and QD under the same conditions are shown in (c) and
(d), respectively. Dashed lines indicate the positions of the peaks
and separate the system into four states: I, II, SL-I, and SL-II. }
\end{figure}

\begin{figure}[t]
\includegraphics[width=1.0\columnwidth]{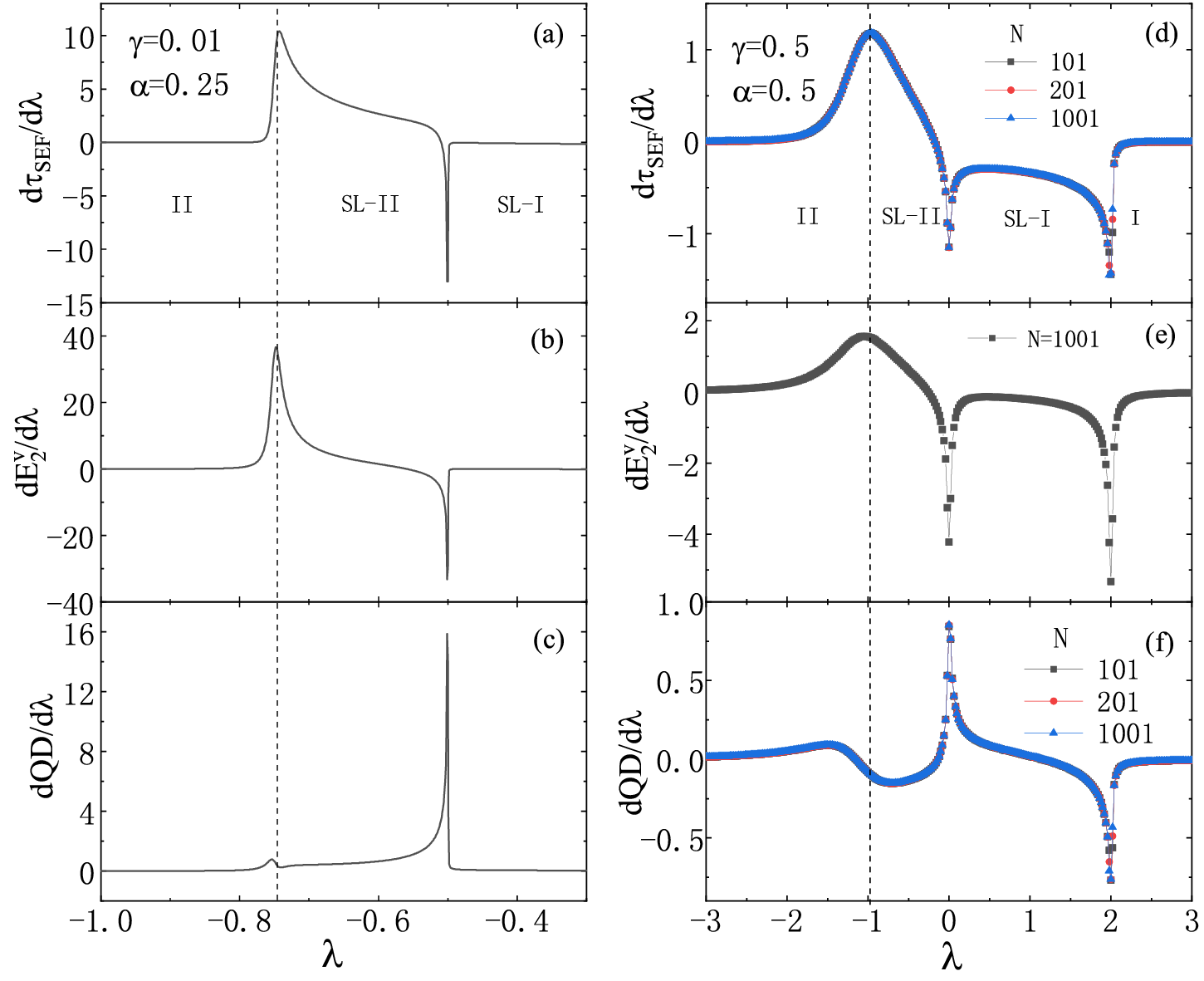}
\caption{\label{fig6} (Color online) The derivative behaviors of
$\tau_{SEF}$ , $E_2^v$ , and $QD$ as a function of $\lambda$. The
left panel represents the case where $\gamma=0.01$, $\alpha=0.25$,
and $\beta=0$ with $N=1001$, while the right panel represents the
case where $\gamma=\alpha=0.5$ and $\beta=0$ with different values
of $N$. The position of the peak in $dE_2^v/d\lambda$ is indicated
by the dashed line, which marks the phase transition point from
SL-II to IV.}
\end{figure}

Using the Jordan-Wigner, Fourier, and Bogoliubov transformations
successively~\cite{Sachdev99}, the spin model can be mapped into a
fermion model, and then the following diagonalized analytical
solution can be obtained in momentum space:
$H=\sum_{k}2\varepsilon_{k}(d_{k}^{+}d_{k}-\frac{1}{2})$~\cite{Titvinidze03},
where ${\varepsilon_k=\sqrt{\epsilon_k^{2}+[\gamma \sin
(x_k)]^{2}}}$ with $\epsilon_k=\lambda
-\cos(x_k)-2\alpha\cos(2x_k)-2\beta\cos(3x_k)$ and $x_k=2\pi k/N$ is
the energy spectrum(Ref.~\cite{LYC11}). According to
Ref.~\cite{Gu08,Wang04}, the reduced density matrix for two
arbitrary sites $i$ and $j$ at arbitrary distance $r=j-i$ can be
obtained as follows
\begin{align}\label{eq:11}
\rho_{r}= \rho_{ij}=\left(
  \begin{array}{cccc}
    u^+ & 0 & 0 & y^- \\
    0 & z & y^+ & 0 \\
    0 & y^+ & z & 0 \\
    y^- & 0 & 0 & u^- \\
  \end{array}
\right)
\end{align}
where the elements can be calculated from the correlation functions
\begin{eqnarray}\label{eq:12}
&&u^{\pm}=\frac{1}{4}(1\pm2\langle\sigma^z\rangle+\langle\sigma_0^z\sigma_r^z\rangle),\nonumber\\
&&z=\frac{1}{4}(1-\langle\sigma_0^z\sigma_r^z\rangle),\nonumber\\
&&y^\pm=\frac{1}{4}(\langle\sigma_0^x\sigma_r^x\rangle\pm\langle\sigma_0^y\sigma_r^y\rangle)
\end{eqnarray}

The mean magnetization and correlation functions can be written
as~\cite{Barouch70,Barouch71}
\begin{align}\label{eq:13}
&\langle\sigma^z\rangle=\frac{1}{N}\sum_k\frac{\epsilon_k}{\varepsilon_k},\nonumber\\
&\langle\sigma_0^x\sigma_r^x\rangle=\left|\begin{array}{cccc}
                                             a_{-1} & a_{-2} & \cdots & a_{-r} \\
                                             a_{0} & a_{-1} & \cdots & a_{-r+1} \\
                                             \vdots & \vdots & \vdots & \vdots \\
                                             a_{r-2} & a_{r-3} & \cdots & a_{-1} \\
                                           \end{array}\right|,\nonumber\\
&\langle\sigma_0^y\sigma_r^y\rangle=\left|\begin{array}{cccc}
                                             a_{1} & a_{0} & \cdots & a_{-r+2} \\
                                             a_{2} & a_{1} & \cdots & a_{-r+3} \\
                                             \vdots & \vdots & \vdots & \vdots \\
                                             a_{r} & a_{r-1} & \cdots & a_{1} \\
                                           \end{array}\right|,\nonumber\\
&\langle\sigma_0^z\sigma_r^z\rangle=\langle\sigma^z\rangle^2-(a_ra_{-r}),
\end{align}
where ${a_r=-\sum_k[\cos(x_kr)\epsilon_k+\gamma\sin(x_kr)\sin(x_k)]}
/(N\varepsilon_k)$. Then the reduced density matrices for a
single-site $\rho_i$ or two-site $\rho_{ij}$ can be obtained, and
all the detectors introduced in Sec.~\ref{sec:level2} can be
calculated.

\begin{figure}[t]
\includegraphics[width=0.8\columnwidth]{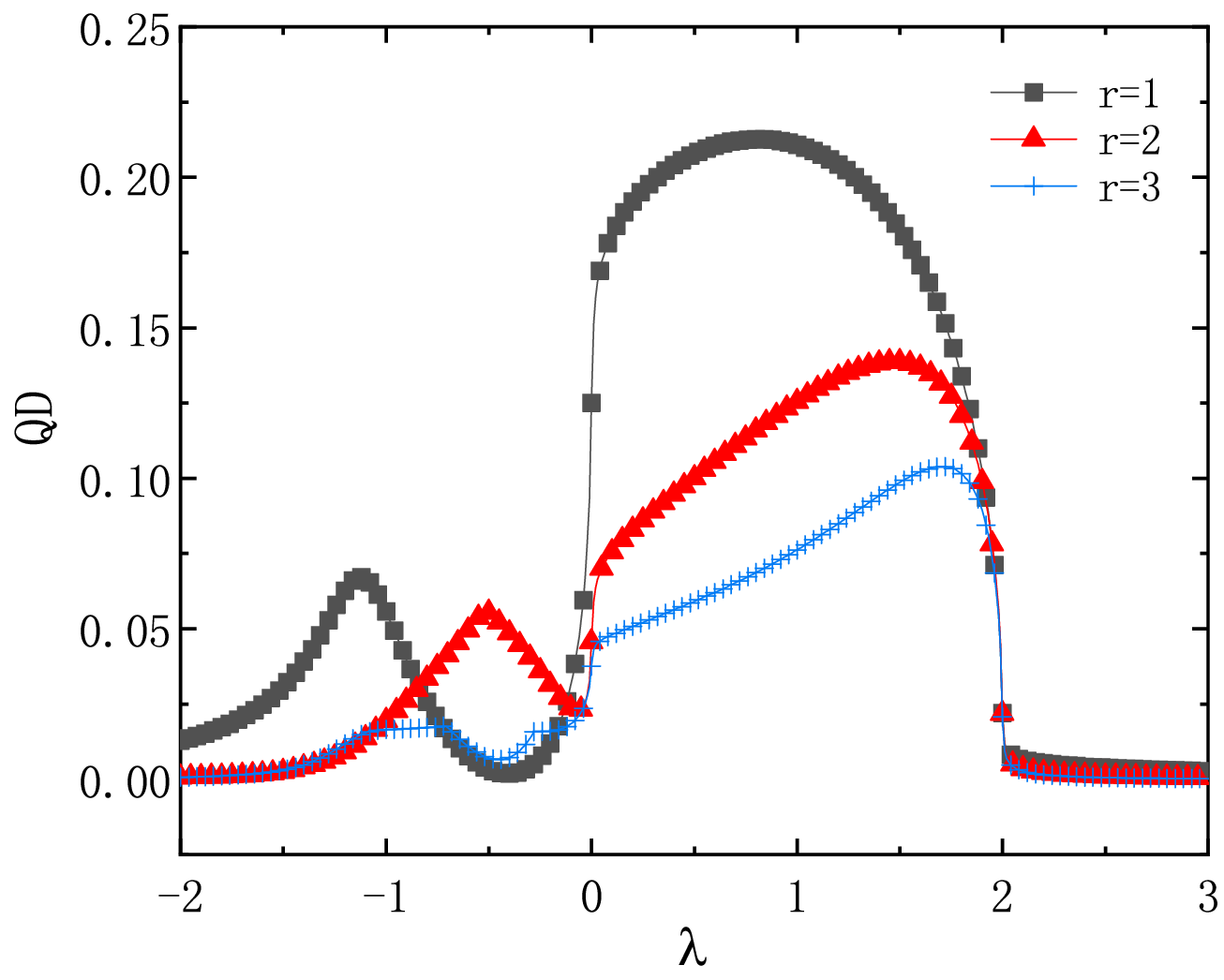}
\caption{\label{fig7} (Color online) QD as a function of $\lambda$
for different $r$ at $\gamma=\alpha=0.5$, $\beta=0$, and $N=1001$.
There exists singularities at a non-phase transition point when
$\lambda<0$ for $r>1$ cases.}
\end{figure}

We first consider the case of $\beta=0$, where the phase diagram is
known~\cite{LYC11,Hu20}. When $|\lambda|$ is large enough, all spins
become polarized, and the system enters a spin-saturated phase.
However, when $|\lambda|$ is small, there is only one spin liquid
phase for small $\alpha$ values, and the phase transition between it
and the spin-saturated phase belongs to the Ising-like
order-disorder type QPT. As $\alpha$ increases, another type of spin
liquid phase appears in the system~\cite{Titvinidze03,LYC11,Hu20}.
The results of $\tau_{SEF}$, QD, and EOF as a function of $\lambda$
for $\gamma=0$ and $\alpha=0.5$ are plotted in the left panel of
Fig.~\ref{fig5}. One can see that there are three obvious turning
points in both the curves of $\tau_{SEF}$ and QD. Obviously, it
corresponds to the spin-polarized state in the region where
$\lambda<-1.12$ (II) and $\lambda>2$ (I). Because the magnetic field
in these regions is strong enough, all spins become polarized, and
the entanglement and correlation between spins disappear. As a
result, the values of $\tau_{SEF}$ and QD are zero in this phase.
The turning point at $\lambda=0$ separates the two spin liquid
phases, SL-I and SL-II. Here, $\tau_{SEF}$ and QD demonstrate
equivalent capabilities in detecting quantum phase transitions,
which is more evident in their derivative behavior (see
Fig.~\ref{fig5}(c) and Fig.~\ref{fig5}(d)), where the non-analytical
peak positions, which reflect QPTs, are entirely consistent. As a
comparison, when $ \lambda<0 $, EOF is always zero, making it
impossible to detect the QPT located at $\lambda=-1.12$(see
Fig.~\ref{fig5}(b)). At the same time, the peak of $dQD/d\lambda$
that reflects the occurrence of this QPT is also very weak.
Therefore, it seems that $\tau_{SEF}$ is a more reliable indicator
for detecting the QPTs in the system. Further study has confirmed
this point.

In Fig.~\ref{fig6}, we present the results of the derivatives of
$\tau_{SEF}$ and QD for various values of $\gamma$ and $\alpha$. The
two-site entropy $E_2^v$ is also calculated for comparison. When
$\gamma=0.01$ and $\alpha=0.25$ (left panel in Fig.~\ref{fig6}),
three detectors provide a completely consistent result for the
position of the CP that separates the SL-I and SL-II states.
However, during the quantum phase transition from the SL-II state to
the spin-polarized II state, there is a noticeable peak behavior in
$d\tau_{SEF}/d\lambda$, and the location of the reflected phase
transition, $\lambda_m\approx-0.747$, closely matches the result
obtained from $dE_2^v/d\lambda$. In contrast, only a small, less
noticeable bump is shown here for $dQD/d\lambda$, and its location
has an obvious deviation from $\lambda_m$.

\begin{figure}[t]
\includegraphics[width=0.9\columnwidth]{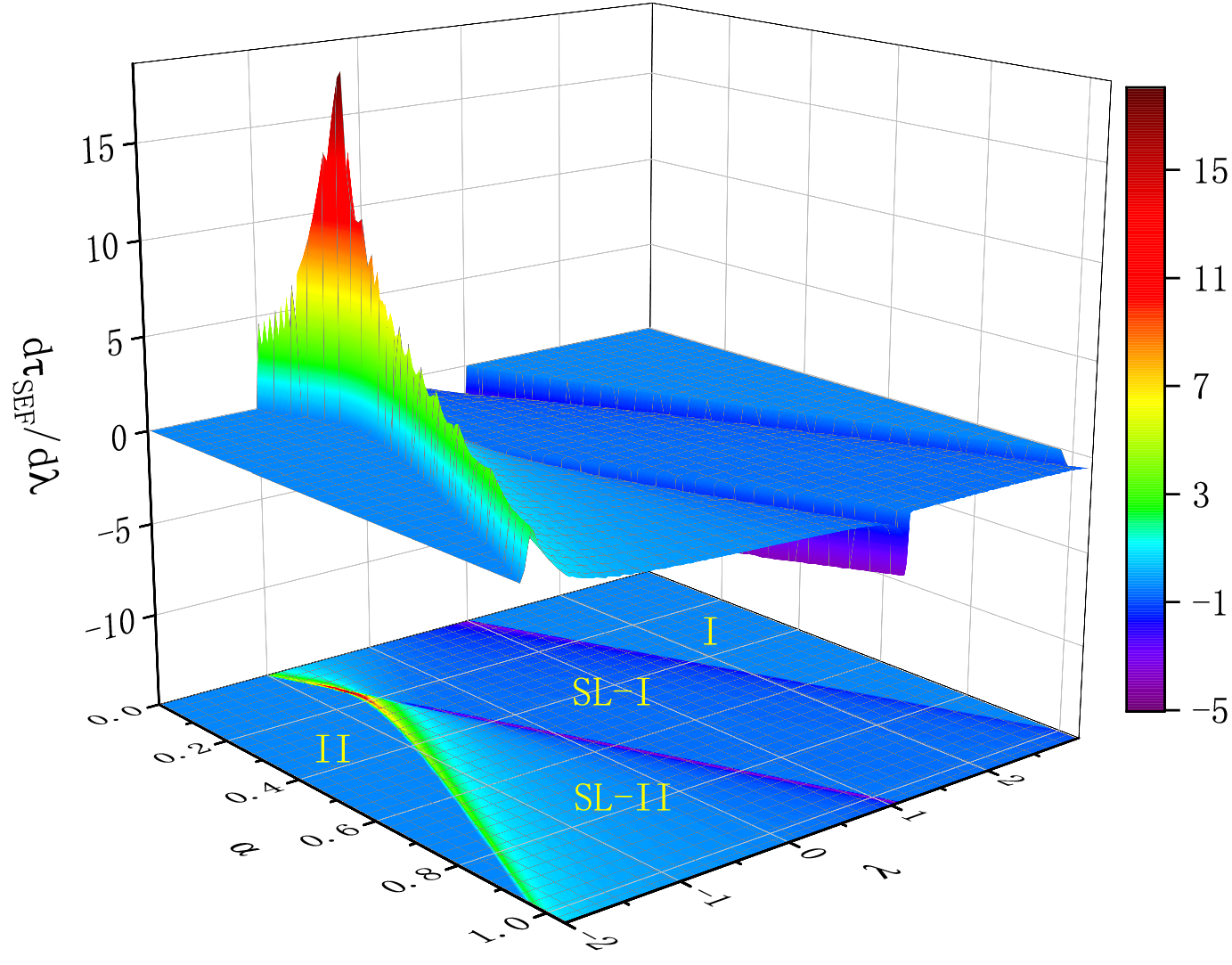}
\caption{\label{fig8} (Color online) The derivative
$d\tau_{SEF}/d\lambda$ and its contour map as functions of $\alpha$
and $\lambda$ for $\gamma=0$, $\beta=0$ and $N=5001$. }
\end{figure}

The comparison of the results for $\gamma=\alpha=0.5$ is even more
pronounced (see the right panel of Fig.~\ref{fig6}). The two CPs
indicated by the three detectors, $\lambda_{c_1}\approx2$ and
$\lambda_{c_2}\approx0$, which separate the SL-I state from the
spin-polarized state I and the SL-II state, are completely
consistent. However, at the third CP $\lambda_{c_3}$, $dQD/d\lambda$
yields a significantly different result compared to the other two
detectors. The peak positions given by $d\tau_{SEF}/d\lambda$ and
$dE_2^v/d\lambda$ that reflect the occurrence of the phase
transition at $\lambda_{c_3}=-1$ are basically consistent, while
$dQD/d\lambda$ has no obvious singularity in this area, and its slow
peak position also differs significantly from the results for
$d\tau_{SEF}/d\lambda$. Moreover, there is no finite-size effect;
all the results are less affected by the system sizes, as shown in
Figs.~\ref{fig6}(d) and (e), where all the curves for different $N$
almost collapse into a single one.

When it comes to the QD itself (see Fig.~\ref{fig7}), except for the
dramatically changing points at $\lambda\approx2$ and
$\lambda\approx0$, which correspond to the CPs $\lambda_{c_1}$ and
$\lambda_{c_2}$, respectively, there is a peak behavior at
$\lambda=-1.13$ for the case when $r=1$. However, the peak position
deviates significantly from the CP $\lambda_{c_3}$, and there are
other non-analyzed points at significantly different $\lambda$ for
cases where $r>1$. This is consistent with the conclusion that the
singularity of QD does not always correspond to a QPT~\cite{Yu16}.
The possible reason is that the calculations of QD involve the min
function procedures in Eq.(\ref{eq:4}), which may destroy the
analyticity of elements in the density matrix~\cite{SZY10}.
Therefore, compared to QD, multipartite quantum entanglement has a
clear advantage in reflecting the phase transition here.

\begin{figure}[t]
\includegraphics[width=0.9\columnwidth]{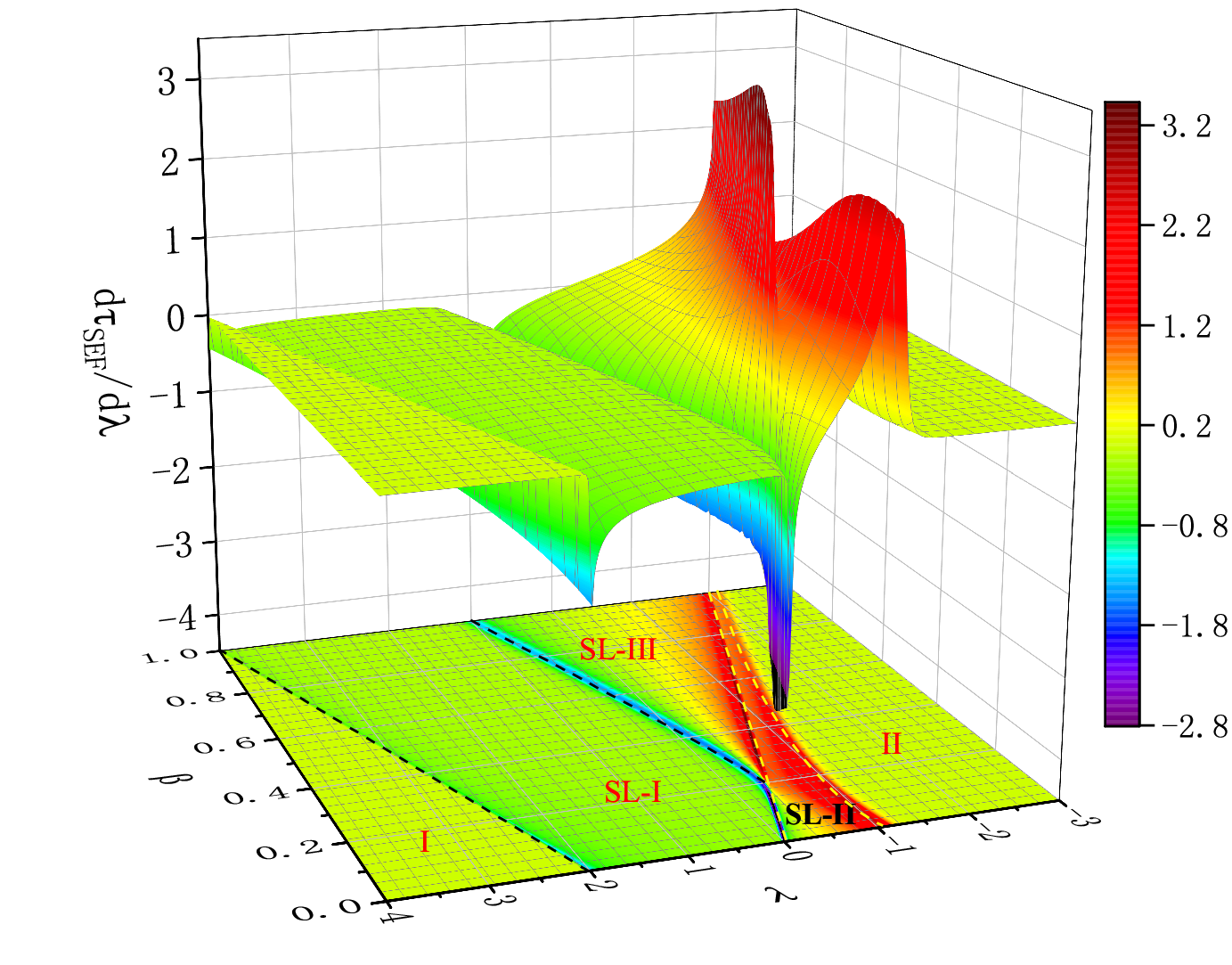}
\caption{\label{fig9}(Color online) The derivative
$d\tau_{SEF}/d\lambda$ and its contour map as functions of $\beta$
and $\lambda$ for $\gamma=0.1$, $\alpha=0.5$, and $N=1001$. The
dashed lines serve as a guide for the points of extremes in the
curves. }
\end{figure}

Using the derivative of multipartite entanglement
$d\tau_{SEF}/d\lambda$, we obtain the phase diagram of the system
for the case when $\beta=0$. The derivatives $d\tau_{SEF}/d\lambda$
and its contour map for $\gamma=0$ as functions of $\lambda$ and
$\alpha$ are shown in Fig.~\ref{fig8}. The sharp peaks and valleys
in the figure clearly separate these three phases, dividing the
system into four regions. Regions I and II correspond to the
spin-polarized state, while regions SL-I and SL-II correspond to two
spin liquid states, respectively. Actually, for a comparison, we
choose the same parameter conditions as that in Ref.~\cite{Hu20}.
The phase diagram given by $d\tau_{SEF}/d\lambda$ are consistent
with those in Refs.~\cite{Titvinidze03,Hu20} given by the steered
quantum correlation, but $\tau_{SEF}$ can avoid the problem of
non-analytic points appearing at non-critical points for the steered
quantum correlation(see details in Ref.~\cite{Hu20}).

We then consider the influence of the four-spin interaction $\beta$
on the QPTs of the system. The results for $d\tau_{SEF}/d\lambda$
with $\gamma=0.1$ and $\alpha=0.5$ are shown in Fig.~\ref{fig9}. The
sharp peaks or sudden drops in the value of the derivative of the
multipartite entanglement clearly indicate the regions of
criticality. Except for the already known quantum states I, II,
SL-I, and SL-II, which can be easily recognized by their continuous
existence from the $\beta=0$ case, a new quantum state named SL-III
appears when $\beta>0.2$. As $\beta$ increases, the SL-III phase
expands to a larger parameter range, while the region where the
SL-II phase exists gradually shrinks. However, the parameter range
where phase SL-I exists is almost unaffected. This indicates that
phases SL-II and SL-III are related to the multi-site interactions
in the system, and $\beta$ is beneficial for the stability of
quantum phase SL-III. Although the specific physical properties of
the new state need to be further studied in the future, the results
of multipartite entanglement provide clear evidence of its
existence.

\section{\label{Sec:level7} Summary}
To summarize, we have examined the efficacy and superiority of the
residual entanglement $\tau_{SEF}$ as a measure of multipartite
entanglement in determining quantum phase transitions in various
typical physical systems. After comparing several commonly used
detectors known for their superior performance in determining
quantum phase transitions, it was discovered that multipartite
quantum entanglement possesses distinct advantages.

The bipartite quantum entanglement $E_2^v$ cannot characterize the
BKT-type QPT in the XXZ model. To accurately detect the location of
the topological QPT in the SSH model, apart from the multipartite
entanglement $\tau_{SEF}$, other bipartite detectors require the use
of derivative behavior and size scaling analysis. For the EOF, the
value of some quantum states in the XYMI model will become zero,
which cannot reflect the occurrence of phase transitions. The
phenomenon of QD in characterizing the SL-II to spin-polarized II
phase transition is not readily apparent and it may exhibit
singularity at a non-phase transition point. In contrast, the
residual entanglement $\tau_{SEF}$ completely avoids the issues of
these detectors and can unequivocally identify the occurrence of
these quantum phase transitions.

We believe that the mechanism is that the residual entanglement
$\tau_{SEF}$  measures the multipartite entanglement in the system,
which incorporates entanglement information concerning both the
considered spin and all other spins. This allows it to provide an
overall and comprehensive reflection of changes in the entanglement
degrees of the system. In contrast, detectors based on bipartite
correlation, which reflect the entanglement or correlation
information between partial states, are easily affected by the
chosen lattice site positions and are not immune to the influence of
interactional structures. This is especially true for models such as
the SSH and XYMI models, where asymmetry and multi-spin interactions
are present. The localized nature of these detectors limits their
ability to reflect overall information changes in the system. Hence,
multipartite quantum entanglement possesses inherent advantages over
bipartite entanglement and correlation in determining quantum phase
transitions. Following this line of thought, and considering that
other measures of quantum correlations may contain more quantum
correlation information than quantum entanglement, we believe that
multipartite quantum correlations may have more advantages in
characterizing quantum phase transitions. Further research in this
area is needed in the future. In addition, experimentally, the
multipartite entanglement based on residual entanglement of
formation can be obtain via quantum state
tomography~\cite{James2001} on one-site and two-site reduced density
matrices of the ground state, which can be used to calculate the
relevant bipartite entanglements. We hope that this paper offers
valuable theoretical insights for future experimental efforts in
this field.

Furthermore, by utilizing the derivative of $\tau_{SEF}$, we present
the quantum phase diagram of the XYMI model and reveal a new spin
liquid phase that emerges from the interaction of four spins.

\begin{acknowledgments}
We acknowledge financial support from the National Natural Science
Foundation of China (Grants No. 12074376, No. 52072365, and
No.11575051), the Beijing Municipal Natural Science Foundation
(Grants No.1222027), the NSAF (Grants No. U1930402), the Heibei
NSF(Grant No. A2021205020), the Fundamental Research Funds for the
Central Universities, and the International Partnership Program
(Grant No. 211211KYSB20210007) of the Chinese Academy of Sciences.

\end{acknowledgments}

\balance

\clearpage

\end{document}